\documentclass[preprint,12pt]{elsarticle}

\usepackage{amssymb}
\usepackage{amsmath}
\usepackage{lineno}

\journal{Chaos, Solitons \& Fractals}

\begin{document}

\begin{frontmatter}

\title{Evolution of cooperation among migrating resource-oriented agents under environmental variability}

\author[a1]{Masaaki Inaba\corref{cor1}}
\author[a2]{Eizo Akiyama}

\cortext[cor1]{Corresponding author. Email: masaaki.inaba@gmail.com}

\affiliation[a1]{%
  organization={Graduate School of Science and Technology, University of Tsukuba},
  addressline={1-1-1 Tennodai},
  city={Tsukuba},
  postcode={305-8577},
  state={Ibaraki},
  country={Japan}}

\affiliation[a2]{%
  organization={Institute of Systems and Information Engineering, University of Tsukuba},
  addressline={1-1-1 Tennodai},
  city={Tsukuba},
  postcode={305-8577},
  state={Ibaraki},
  country={Japan}}

\begin{highlights}
\item Modeling environmental variability as a driver of Middle Stone Age human cooperation
\item Environmental variability and agent mobility can promote the evolution of cooperation
\item Environmental variability can disrupt stable defector groups in resource-rich areas
\item Agent interactions and mobility can foster cooperative group formation
\item Cooperation can evolve without requiring enhanced cognitive abilities
\end{highlights}

\begin{abstract}
Cooperation is fundamental to human societies.
While several basic theoretical mechanisms underlying its evolution have been established, research addressing more realistic settings remains underdeveloped.
Drawing on the hypothesis that intensified environmental fluctuations drove early behavioral evolution in humans during the Middle Stone Age in Africa, we examine the effects of environmental variability and human mobility on the evolution of cooperation.
In our model, the variability is represented by randomly moving resource-rich spots across a two-dimensional space, and the mobility is represented by resource-seeking migration of agents.
These agents interact cooperatively or competitively for resources while adopting behavioral strategies from more successful neighbors.
Through extensive simulations of this model, we reveal three key findings:
(i) with sufficient agent mobility, even modest environmental variability promotes cooperation, but further variability does not enhance cooperation;
(ii) with any level of environmental variability, agent mobility promotes cooperation; and
(iii) these effects occur because the joint effect of environmental variability and agent mobility disrupts defector groups in resource-rich areas, forming cooperator groups on those sites.
Although previous studies examine environmental variability and mobility separately, to our knowledge this is the first study to analyze their joint effects on the evolution of cooperation.
These findings suggest that environmental variability can promote cooperative group formation without enhanced cognitive abilities, providing new insights into the evolution of human cooperation and, by extension, sociality.
\end{abstract}


\end{frontmatter}

% \linenumbers

\section{Introduction}\label{sec:introduction}

Cooperative behavior is a fundamental feature of human societies.
Yet this widespread behavior poses an evolutionary paradox:
while evolutionary theory suggests that individuals act to maximize their own fitness, cooperation appears to favor the benefit of others or groups over that of individuals.
A substantial body of research \cite{Nowak2006, West2007, Perc2010, Rand2013, Perc2017, West2021} has addressed this puzzle, identifying various theoretical mechanisms that explain the evolution of cooperation.
However, several influential factors on cooperation in natural settings---such as environmental variability, mobility, complex networks, norms, and memory---remain theoretically underexplored.
Here, we focus on how environmental variability and agent mobility shape the evolutionary dynamics of cooperation.

Although most theoretical studies on the evolution of cooperation assume static environmental conditions, some studies \cite{Brockhurst2007, Assaf2013, Miller2015, Amirjanov2017, Stojkoski2021, Kroumi2021} have examined cooperation under variable environmental conditions.
While these studies have primarily focused on biological and physical contexts, we approach the study of cooperation under environmental variability from an anthropological perspective, specifically drawing on the Variability Selection Hypothesis (VSH) \cite{Potts1998, Potts2015, Potts2018, Lupien2021}.
The VSH posits that variable and unpredictable environments during the Middle Stone Age (MSA) in Africa were crucial drivers of human evolution.
This hypothesis is based on the temporal correspondence between periods of increased environmental variability and the sophistication of archaeological artifacts, though the underlying causal dynamics are not yet fully resolved.
Our previous work \cite{Inaba2025} contributes to clarifying these dynamics by demonstrating that environmental variability can promote cooperation through the interplay between environmental disruptions and social network structures that support cooperation.
However, this study does not incorporate mobility---an essential trait of pre-sedentary MSA human populations.

The evolution of cooperation among mobile agents has been actively studied.
Dugatkin and Wilson (1991) \cite{Dugatkin1991} conducted pioneering research in this area, and Vainstein et al. (2007) \cite{Vainstein2007} later established that mobility promotes cooperation under very minimal assumptions.
More recently, various migration strategies \cite{Cong2012, Chen2012, He2020, Dhakal2020, Ren2021, Yang2023, Zhang2025} have been proposed that differ in the trigger and distance of migration, and the evolution of cooperation under each strategy has been investigated.
Nevertheless, in the area of the evolution of cooperation with migration, research addressing exogenous environmental variability has yet to be undertaken.

In summary, although research has separately examined cooperation under environmental variability and cooperation with mobility, no studies have integrated both factors.
Motivated by our interest in environmental variability and human behavioral evolution during the MSA in Africa, we investigate how environmental variability and agent mobility jointly affect the evolution of cooperation.

\section{Model}\label{sec:model}

We develop a multi-agent simulation model to investigate how agent mobility and environmental variability jointly influence the evolution of cooperation in spatially structured populations.
Due to limited archaeological data on spatial resource distributions and hominin behavioral patterns during the MSA, we adopt a deliberately abstracted approach that prioritizes identifying fundamental mechanisms over reproducing specific historical scenarios.
The model incorporates four key processes: (i) environmental variability on a two-dimensional lattice, (ii) pairwise game interactions, (iii) conditional agent migration driven by resource availability, and (iv) strategy updating.

\subsection{Environmental variability}\label{sec:ev}

The spatial structure is represented by a two-dimensional lattice with periodic boundary conditions.
$N$ agents are randomly distributed across cells, with each cell containing at most one agent.
Each cell maintains a resource level that varies both spatially and temporally.
We define a Source of Resources (SoR) as a focal point that generates spatial resource gradients in the environment, serving as a simplified representation of natural foraging areas commonly found near rivers, lakes, or coastlines.
Each SoR creates a gradient where resource availability typically decreases with distance from it.
Multiple SoRs may coexist, forming overlapping zones of resource abundance.

In this model, each agent accumulates resources (represented by its fitness value) through interactions, and local prosperity is defined by a resource threshold $\theta_{x,y}$ for each cell $(x, y)$, where $0 \leq \theta_{x,y} \leq 1$.
Agents with fitness below the threshold $\theta_{x,y}$ are more likely to migrate and update strategies, while those above the threshold remain unchanged.
Thus, locations with lower thresholds impose less pressure for behavioral change, indicating that the locations are resource-rich and prosperous.
The threshold $\theta_{x,y}$ is determined by the cumulative influence of all SoRs and is calculated as:
\begin{equation}
\theta_{x,y} = \frac{D_{x,y} - D_{\min}}{D_{\max} - D_{\min}}, \quad D_{x,y} = \sum_{i=1}^{n} d_{x,y}^{(i)}
\end{equation}
where $d_{x,y}^{(i)}$ denotes the Euclidean distance from cell $(x, y)$ to the $i$-th SoR, calculated under periodic boundary conditions, and $D_{\min}$ and $D_{\max}$ are the minimum and maximum values of $D_{x,y}$ across the entire grid, respectively.

\begin{figure}[!ht]
\centering
\includegraphics[width=1.0\linewidth]{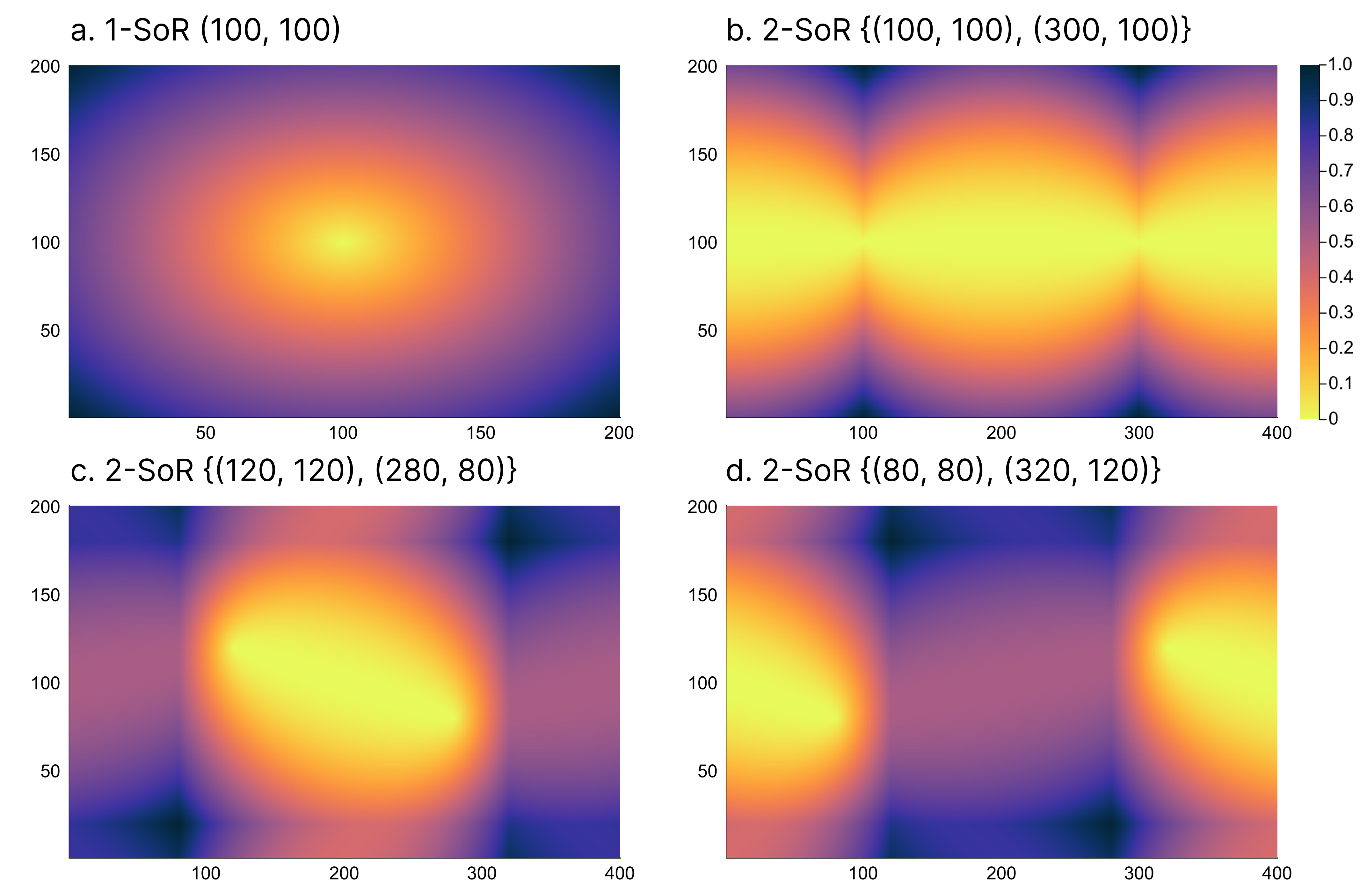}
\caption{
Spatially heterogeneous prosperity patterns generated by SoR(s).
Colors indicate the resource threshold $\theta_{x,y}$.
(a) A single SoR located at (100, 100) on a $200\times200$ grid generates a concentric resource gradient.
(b) Two SoRs located at (100, 100) and (300, 100) on a $400\times200$ grid generate a band-shaped resource gradient.
(c) and (d) show examples of SoR shifts over time.
}\label{fig:ev}
\end{figure}

We examine two spatial configurations: 1-SoR and 2-SoR.
In the 1-SoR configuration ($200\times200$ grid), a single SoR generates a concentric resource gradient, capturing the essential geographical pattern of oasis-like environments (Figure~\ref{fig:ev}a).
In the 2-SoR configuration ($400\times200$ grid), two SoRs generate a corridor of resource gradient, capturing the essential geographical pattern of riverine or coastal environments where resources are distributed along a line (Figure~\ref{fig:ev}b).

Environmental variability is modeled through the stochastic movement of SoRs, reflecting the unpredictable nature of landscape dynamics observed during the MSA in Africa.
Each SoR moves with probability $p_{EV} \, (0 \leq p_{EV} \leq 1)$ to a randomly selected adjacent cell within the Moore neighborhood at each time step, with the direction chosen uniformly at random from the eight neighboring cells.
The parameter $p_{EV}$ governs the intensity of environmental variability: $p_{EV} = 0$ corresponds to a static environment, while higher values indicate more intense environmental variability.

Throughout this study, the neighborhood is defined as the Moore neighborhood (eight neighboring cells).
This choice better approximates the continuous spatial movement of SoRs and mobile agents compared to the von Neumann neighborhood (four orthogonal cells), which restricts movement to cardinal directions only.
Additionally, for distance calculations, particularly in determining resource gradients from SoRs, Euclidean distance is employed to produce smooth, circular gradients that better represent natural resource distributions than alternative metrics such as Manhattan or Chebyshev distances.

\subsection{Game}\label{sec:game}

Games represent cooperative or competitive interactions between agents that involve the gain or loss of resources.
Each agent holds a strategy, either cooperation ($C$) or defection ($D$), which determines its behavior in interactions.
At each time step, every agent plays pairwise games with all of its neighbors and accumulates a payoff $\pi_j$, which is then converted into fitness $\omega_j$ ($j \in [1,\ldots,N]$, $0 < \omega_j < 1$), representing the agent's resource level.

The payoff matrix of the game is defined as follows:
\begin{equation}
\begin{array}{c|cc}
  & C & D \\
\hline
C & R & S \\
D & T & P
\end{array}
\end{equation}
where $R = 1$, $0 < T < 2$, $-1 < S < 1$, and $P = 0$.
To ensure the robustness of our results across a range of social dilemma contexts, we consider various game structures, including the Prisoner's Dilemma ($T > R > P > S$), Stag Hunt ($R > T > P > S$), and Snowdrift ($T > R > S > P$) games.

The accumulated payoff $\pi_j$ is transformed into fitness $\omega_j$ using a sigmoid function:
\begin{equation}
\omega_j = \frac{1}{1 + \exp(-k(\pi_j - \pi_0))}
\end{equation}
where $k$ determines the steepness of the sigmoid curve and $\pi_0$ sets the baseline payoff at which $\omega_j = 0.5$.
We set $k = 1$ for moderate sensitivity and $\pi_0 = 4.0$ to center the sigmoid around typical payoff values encountered in mixed-strategy populations.
Given that agents can accumulate payoffs from up to eight neighbors, the theoretical payoff range is $8S$ to $8R$ for cooperators and $8P$ to $8T$ for defectors.
With our parameter constraints, this yields a payoff range of $-8 < \pi_j < 16$.

\subsection{Migration}\label{sec:migration}

Agents migrate conditionally, depending on their fitness and the local resource threshold.
If an agent's fitness $\omega_j$ falls below the resource threshold $\theta_{x,y}$ at its current location, it migrates with probability $p_M$ and remains with the complementary probability $1 - p_M$.
The migration direction follows a resource-oriented bias: with probability $p_{SoR}$ (SoR orientation), the agent moves toward a neighboring cell with the lowest $\theta_{x,y}$ value, and with the complementary probability $1 - p_{SoR}$ it moves randomly within the neighborhood.
Cells already occupied by other agents are excluded from the candidate destinations.
Agents migrate asynchronously in a randomized order to avoid movement conflicts.

\subsection{Strategy update}\label{sec:strategy}

At each time step, all agents update their strategies synchronously.
Agents whose fitness $\omega_j$ falls below the local threshold $\theta_{x,y}$ adopt the strategy of the most successful individual (i.e., the one with the highest fitness) among its neighbors.
This process is subject to mutation: with probability $\mu$, the adopted strategy is replaced by the alternative strategy (e.g., from $C$ to $D$, or vice versa).

\subsection{Evaluation}\label{sec:evaluation}

To examine the effects of environmental variability and agent mobility on the evolution of cooperation, we perform simulations under a range of parameter configurations (Table~\ref{tab:params}).

Each configuration is evaluated through $100$ independent trials of $10000$ generations each.
As the primary outcome measure, we compute cooperation rate $\phi_C$, defined as the average proportion of agents employing the $C$ strategy during the final $5000$ generations, averaged over all trials.

\begin{table}[!ht]
\centering
\caption{Model parameters used in the simulations.}
\label{tab:params}
\begin{tabular}{cll}
\hline
\textbf{Parameter} & \textbf{Description} & \textbf{Value options} \\
\hline
$W \times H$ & Grid dimensions & $200 \times 200, 400 \times 200$ \\
$N$ & Total number of agents & $500 \times 2^{\{0, 1, 2, 3, 4, 5\}}$ \\
$\phi_C^0$ & Initial frequency of cooperators & $0, 0.5, 1$ \\
$n_{SoR}$ & Total number of SoRs & 1, 2 \\
$p_{EV}$ & Probability of SoR shift & $0$ to $1$ (step: $0.1$) \\
$R$ & Payoff for mutual cooperation & $1$ \\
$T$ & Payoff for defection against cooperator & $0$ to $2$ (step: $0.1$) \\
$P$ & Payoff for mutual defection & $0$ \\
$S$ & Payoff for cooperation against defector & $-1$ to $1$ (step: $0.1$) \\
$p_M$ & Probability of migration & $0$ to $1$ (step: $0.1$) \\
$p_{SoR}$ & Probability of resource-oriented migration & $0$ to $1$ (step: $0.1$) \\
$\mu$ & Probability of mutation  & $0, 0.01$ \\
\hline
\end{tabular}
\end{table}

\section{Results}\label{sec:results}

\subsection{Influence of environmental variability and agent mobility}\label{sec:key_results}

\begin{figure}[!ht]
\centering
\includegraphics[width=0.7\linewidth]{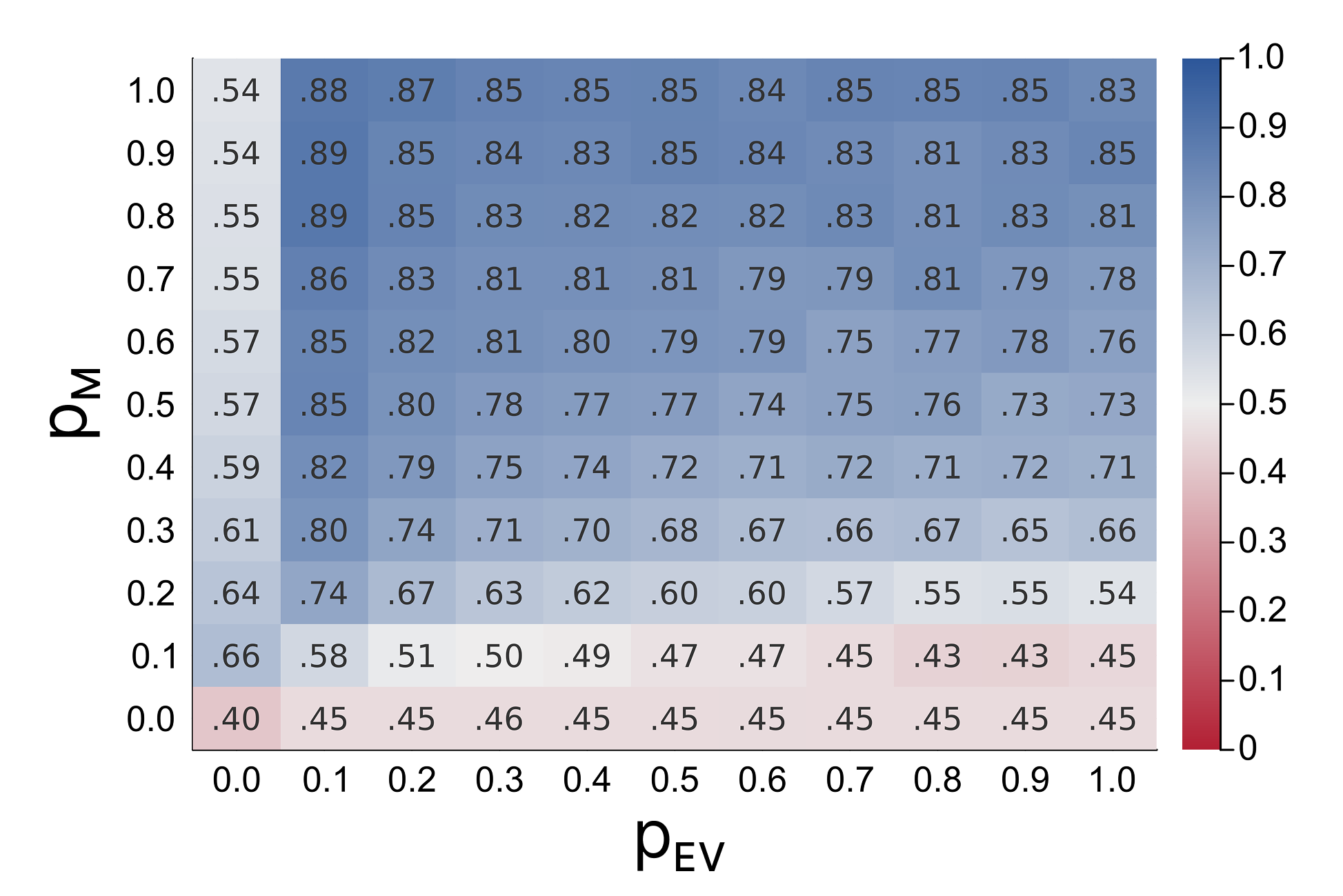}
\caption{
Influence of environmental variability ($p_{EV}$) and agent mobility ($p_M$) on cooperation rate ($\phi_C$).
Each cell shows the mean over 100 independent runs.
Results are shown for a representative parameter set ($N = 1000$, $\phi_C^0 = 0.0$, 2-SoR, $T = 1.2$, $S = -0.2$, $p_{SoR} = 0.1$, $\mu = 0.01$); qualitatively similar patterns are observed for other parameter configurations.
Standard deviations across runs are $< 0.15$ for all cells.
}\label{fig:key_result}
\end{figure}

Figure~\ref{fig:key_result} illustrates the combined effects of environmental variability ($p_{EV}$) and agent mobility ($p_M$) on the cooperation rate $\phi_C$.
In stable environments ($p_{EV} = 0$) or with low mobility ($p_M \lesssim 0.2$), cooperation fails to evolve.
However, with sufficient agent mobility ($p_M \gtrsim 0.2$), even modest environmental variability ($p_{EV} = 0.1$) promotes cooperation, but further variability ($p_{EV} > 0.1$) does not enhance cooperation.
Additionally, with any level of environmental variability ($p_{EV} > 0$), agent mobility promotes cooperation.

\begin{figure}[!ht]
\centering
\includegraphics[width=1.0\linewidth]{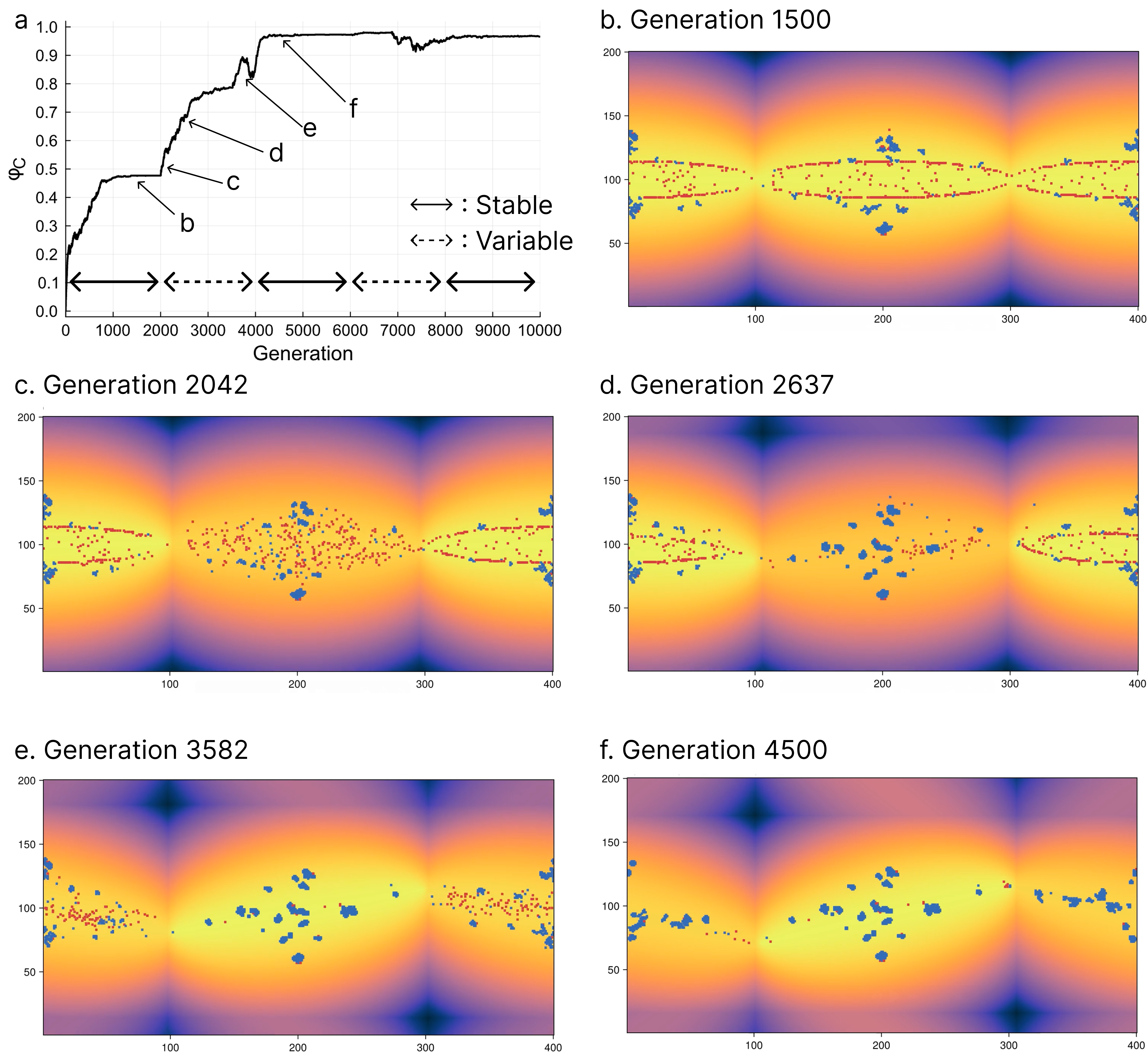}
\caption{
Temporal dynamics under cyclic environmental variability.
Parameter setting: $N = 1000$, $\phi_C^0 = 0.0$, 2-SoR, $T = 1.2$, $S = -0.2$, $p_M = 1.0$, $p_{SoR} = 0.1$, $\mu = 0.01$.
The stable phase corresponds to $p_{EV} = 0$, while the variable phase corresponds to $p_{EV} = 0.1$.
Blue dots represent cooperators; red dots represent defectors.
Background colors indicate the resource threshold $\theta_{x,y}$, as in Figure~\ref{fig:ev}.
}\label{fig:temporal_01_10}
\end{figure}

Figure~\ref{fig:temporal_01_10} further demonstrates the role of environmental variability in the evolution of cooperation.
This representative simulation is presented to elucidate the temporal dynamics underlying the statistical patterns shown in Figure~\ref{fig:key_result}.
Whereas Figure~\ref{fig:key_result} considers fixed levels of variability throughout the simulation, here we apply a cyclic $p_{EV}$ alternating between stable and variable conditions every $2000$ generations, over a total of $10000$ generations.
Under these cyclic conditions, cooperation initially rises to just below $50\%$ during the first stable phase and then plateaus (Figure~\ref{fig:temporal_01_10}a).
A pronounced increase is observed once the system enters the variable phase, reinforcing the conclusion that environmental variability can play a pivotal role in promoting cooperation.

The observed dynamics can be understood as a three-stage process: the formation of a few large defector groups fixed in resource-rich areas; their collapse induced by environmental variability; and the subsequent emergence of several small cooperator groups.
Figures~\ref{fig:temporal_01_10}b--f show snapshots from the simulation in Figure~\ref{fig:temporal_01_10}a.
A full movie is provided in the Supplementary Material.
In the first stable phase (Figure~\ref{fig:temporal_01_10}b), agents in prosperous areas have no need to cooperate or move, whereas those in less prosperous areas must cooperate or move to prosperous areas.
At the boundaries between the two areas, fixed walls are formed by defectors who need neither change their strategies nor move further.
During the next variable phase (Figures~\ref{fig:temporal_01_10}c--e), agents that have been located on boundaries are forced to cooperate or move due to environmental changes, leading to the collapse of the stable defector group (central area of Figure~\ref{fig:temporal_01_10}c).
In place of the defector group, agents form several small cooperator groups to survive even in severe environmental conditions (Figure~\ref{fig:temporal_01_10}d).
Subsequently, the same process occurs in the areas at both ends of the figure (Figures~\ref{fig:temporal_01_10}e,f).

\begin{figure}[!ht]
\centering
\includegraphics[width=1.0\linewidth]{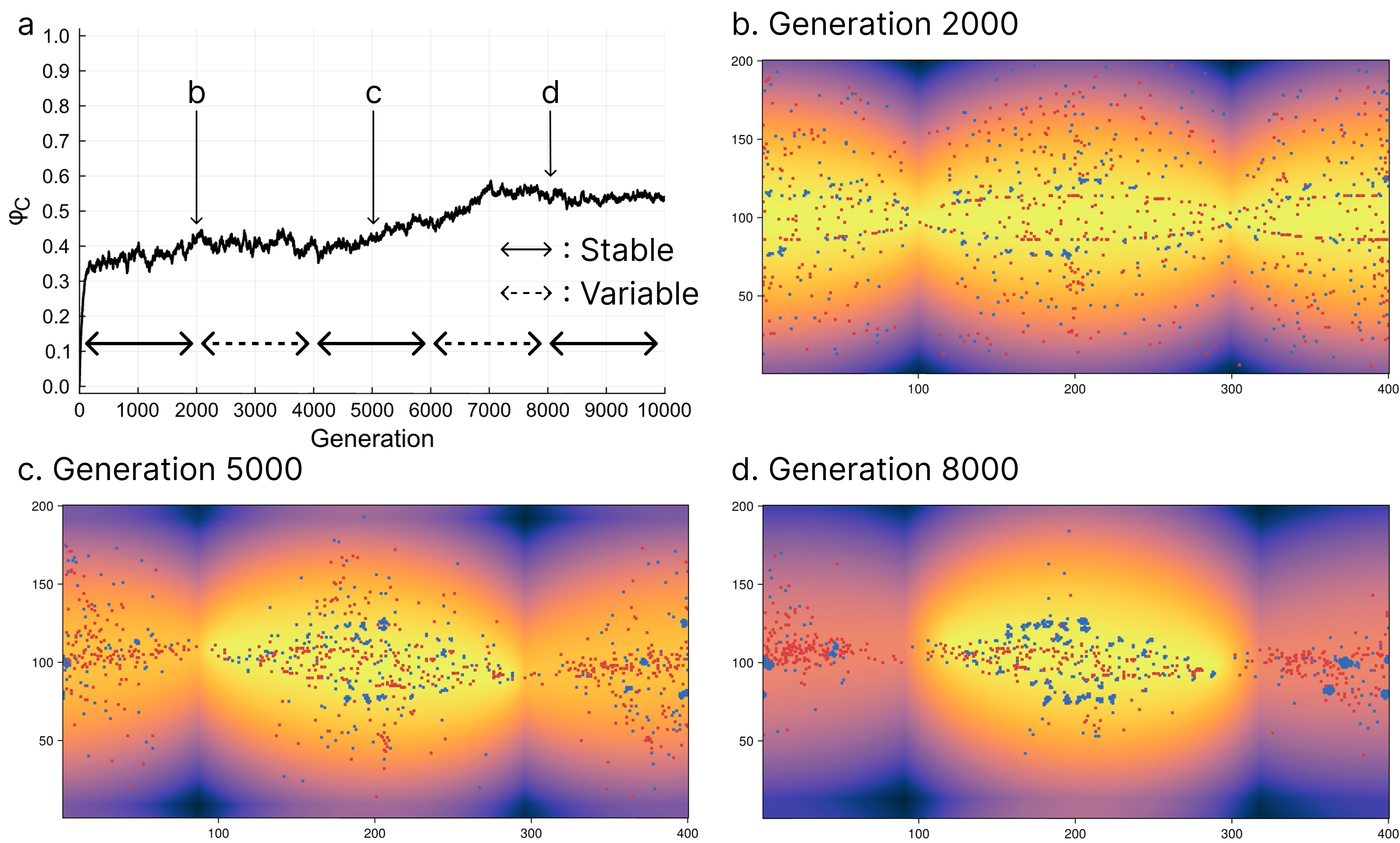}
\caption{
Temporal dynamics under cyclic environmental variability with low agent mobility.
In contrast to Figure~\ref{fig:temporal_01_10} ($p_M = 1.0$), the lower mobility here ($p_M = 0.1$) prevents agents from keeping up with environmental variability.
As a result, cooperation fails to stabilize.
All other parameters and the interpretation of visual elements are the same as in Figure~\ref{fig:temporal_01_10}.
}\label{fig:temporal_01_01}
\end{figure}

On the other hand, if agent mobility is insufficient relative to the intensity of environmental variability, cooperation cannot evolve (see the lower area of Figure~\ref{fig:key_result}).
This occurs because excessively rapid SoR movement prevents the formation of both defector structures and small cooperator groups (Figure~\ref{fig:temporal_01_01}).
Thus, together environmental variability and sufficient agent mobility promote cooperation by preventing fixed defector structures and leading agents to form cooperator groups for survival.

\subsection{Influence of other parameters}\label{sec:other_results}

To gain deeper insights into the findings presented above and assess their robustness, we explored the effects of additional parameters, including population size ($N$), number of SoRs ($n_{SoR}$), initial frequency of cooperators ($\phi_C^0$), SoR orientation ($p_{SoR}$), payoff parameters ($T$, $S$), and mutation rate ($\mu$).

\subsubsection{Population size}

\begin{figure}[!ht]
\centering
\includegraphics[width=0.7\linewidth]{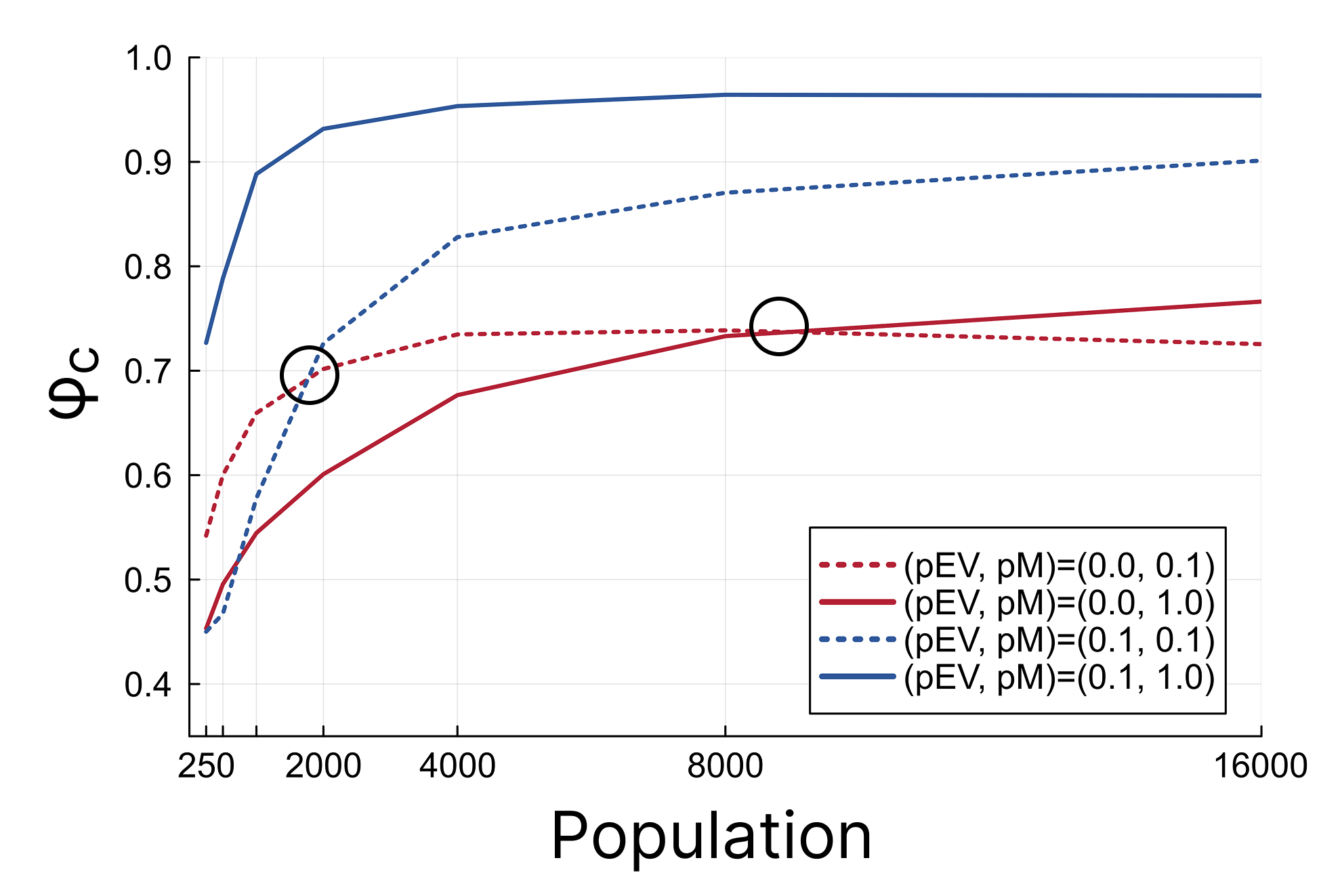}
\caption{
Influence of population size ($N$) on cooperation rate ($\phi_C$).
Parameter setting: $\phi_C^0 = 0.0$, 2-SoR, $T = 1.2$, $S = -0.2$, $p_{SoR} = 0.1$, $\mu = 0.01$.
Results for $N = 32000$ are omitted as they show negligible differences from $N = 16000$.
}\label{fig:population}
\end{figure}

Larger population size promotes cooperation to some extent, as shown in Figure~\ref{fig:population}.
This is because cooperators can more easily find other cooperators when the population is sufficiently large.

Another notable point is that the cooperation rates for $(p_{EV}, p_M) = (0.0, 0.1)$ (the red dashed line) and $(0.1, 0.1)$ (the blue dashed line) exhibit a crossover at around $N = 2000$.
For small populations ($N < 2000$) with low mobility, cooperation evolves more readily without environmental variability.
In stable environments ($(p_{EV}, p_M) = (0.0, 0.1)$), both defector groups and cooperator groups persist once established, although their formation is slow due to low mobility.
In contrast, environmental variability with low mobility ($(p_{EV}, p_M) = (0.1, 0.1)$) creates perpetual fluidity that prevents the formation of both structures.

Larger populations alter this relationship.
Larger populations raise the encounter rate among cooperators.
The higher encounter rate allows cooperator group formation even under environmental variability.
In stable environments, the structures remain fixed, limiting the impact of the higher encounter rate.
In contrast, in fluid environments, the higher encounter rate facilitates group formation.
Therefore, environmental variability becomes advantageous for cooperation above the critical population size.

A less pronounced but similar crossover occurs at around $N = 8000$ between $(p_{EV}, p_M) = (0.0, 0.1)$ (the red dashed line) and $(p_{EV}, p_M) = (0.0, 1.0)$ (the red solid line).
This crossover reflects the spatial constraints of defector groups in stable environments ($p_{EV} = 0$).
For $N \lesssim 8000$, high mobility ($p_M = 1.0$) allows more agents to reach resource-rich areas where they do not need to cooperate.
In contrast, low mobility ($p_M = 0.1$) keeps agents in peripheral areas where they must cooperate to survive.

However, larger populations ($N \gtrsim 8000$) reveal three distinct nested zones:
central rich areas where agents do not need to cooperate, 
surrounding moderate areas where agents can survive through cooperation,
and the most peripheral harsh areas where agents cannot survive even with cooperation because the resource threshold exceeds what cooperation can provide.
Under these conditions, high mobility allows agents to escape from the harsh areas to the moderate cooperative areas, whereas low mobility traps agents in the harsh areas.
Meanwhile, since the central rich areas have already reached their physical capacity due to the large population, further population increases have no effect within these areas.
Thus, the effect of mobility on cooperation rate ($\phi_C$) reverses as population size increases.

\subsubsection{Number of SoRs and initial cooperator rate}

\begin{figure}[!ht]
\centering
\includegraphics[width=1.0\linewidth]{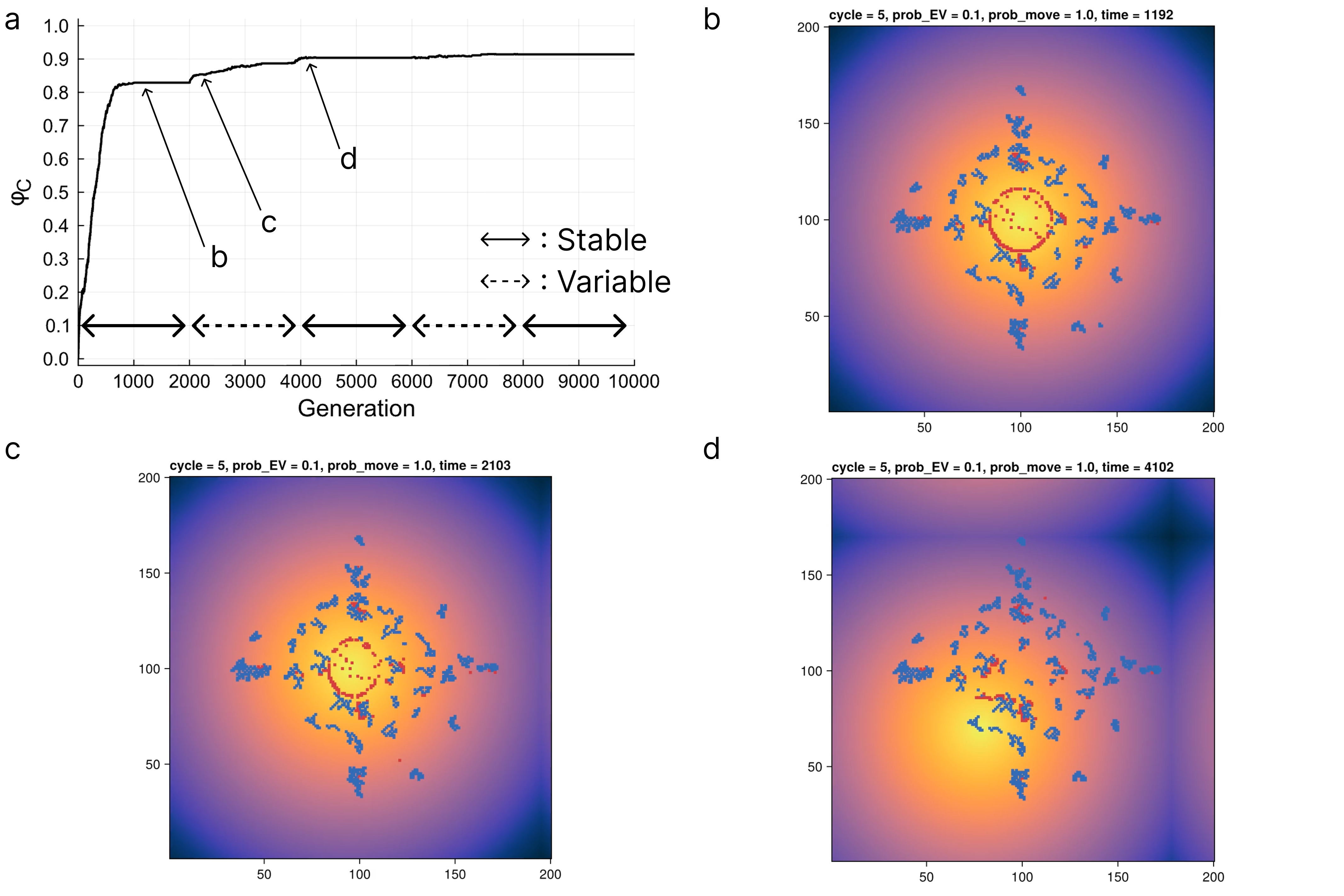}
\caption{
Temporal dynamics under cyclic environmental variability with 1-SoR.
In contrast to Figure~\ref{fig:temporal_01_10} (2-SoR), the impact of the transition from defection to cooperation in the central resource-rich area is less pronounced in the 1-SoR setting.
All other parameters and the interpretation of visual elements are the same as in Figure~\ref{fig:temporal_01_10}.
}\label{fig:1sor}
\end{figure}

\begin{figure}[!ht]
\centering
\includegraphics[width=1.0\linewidth]{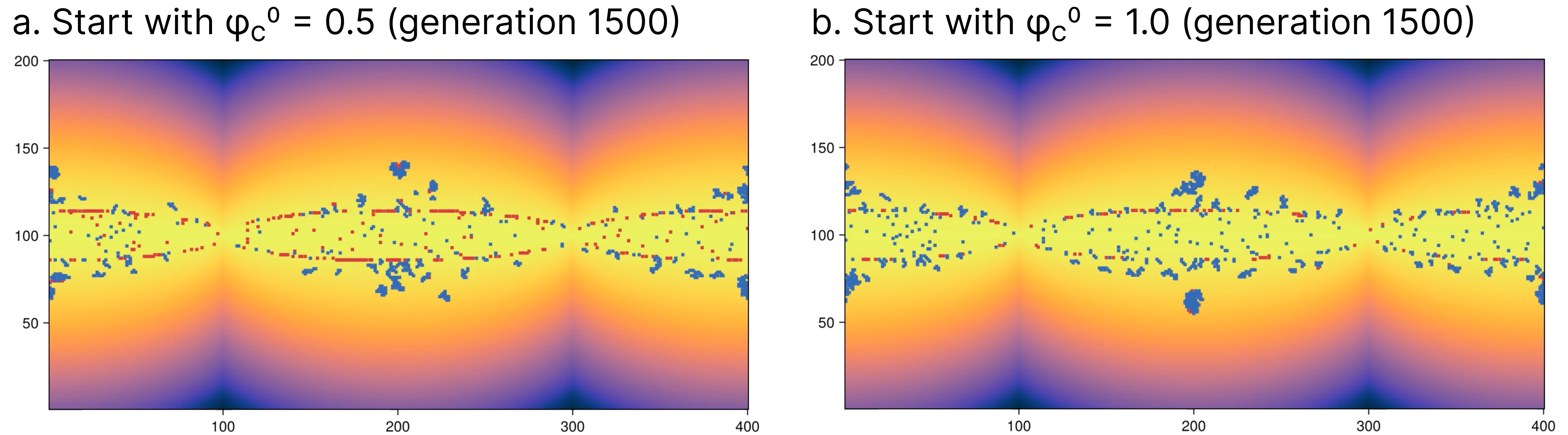}
\caption{
Central resource-rich areas under $\phi_C^0 = 0.5$ or $1.0$.
All parameters except $\phi_C^0$ and the interpretation of visual elements are the same as in Figure~\ref{fig:temporal_01_10}.
}\label{fig:phiC0}
\end{figure}

Both the number of SoRs ($n_{SoR}$) and the initial frequency of cooperators ($\phi_C^0$) have a significant impact on the results.
While the 2-SoR configuration forms large band-shaped defector groups (Figure~\ref{fig:temporal_01_10}b), the 1-SoR configuration forms only a small, circular resource-rich area (Figure~\ref{fig:1sor}b).
When a large defector group collapses and is replaced by small cooperator groups---as observed in the 2-SoR setting---the impact on the overall system is substantial.
In contrast, when a small defector group in the 1-SoR setting undergoes the same replacement, the effect is less conspicuous compared to the 2-SoR case (Figure~\ref{fig:1sor}a).
Regarding $\phi_C^0$, starting with no cooperators ($\phi_C^0 = 0$) in the 2-SoR configuration leads to large defector groups, whereas intermediate or full cooperation ($\phi_C^0 = 0.5$ or $1$) maintains the large structure but with higher cooperator frequency within it (Figure~\ref{fig:phiC0}).
Consequently, these conditions also mask the significant effects of defector group collapse and cooperator group formation observed in the baseline scenario.

We also confirmed that changing the distance metric from Euclidean to Chebyshev or Manhattan alters the size and shape of groups, thereby affecting the results.
However, these effects are primarily attributable to differences in group size rather than shape.
Therefore, comparisons among these distance metrics can be interpreted as theoretically equivalent to our comparison between the 1-SoR and 2-SoR configurations.
In essence, the formation of large defector groups in stable environments is the critical prerequisite for the results described in Section~\ref{sec:key_results}.

\subsubsection{SoR orientation}

\begin{figure}[!ht]
\centering
\includegraphics[width=0.6\linewidth]{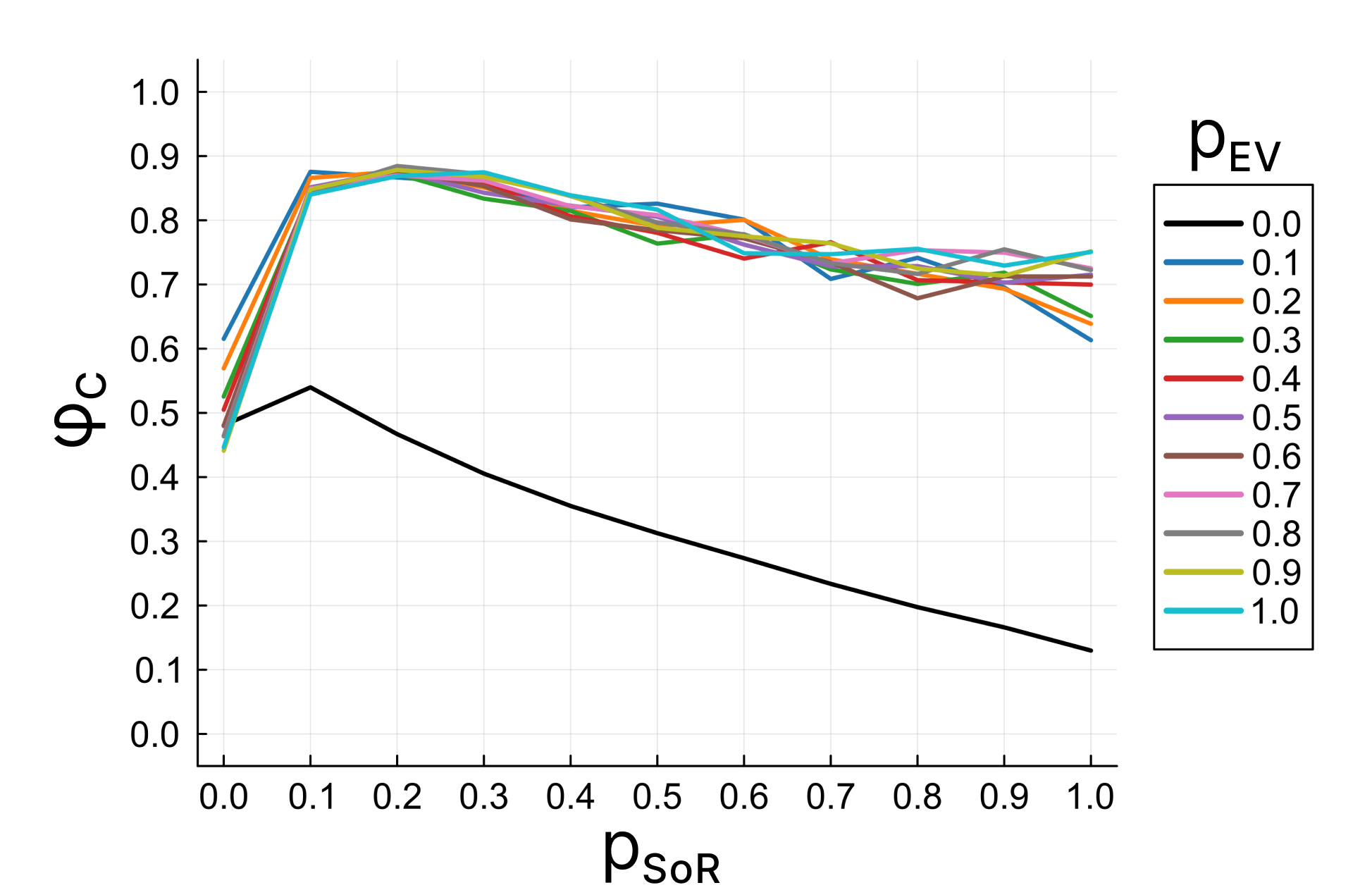}
\caption{
Influence of SoR orientation ($p_{SoR}$) on cooperation rate ($\phi_C$).
$N = 1000$, $\phi_C^0 = 0.0$, 2-SoR, $p_M = 1.0$, $T = 1.2$, $S = -0.2$, $\mu = 0.01$.
}\label{fig:sor_orientation}
\end{figure}

SoR orientation ($p_{SoR}$) noticeably influences cooperation levels (Figure~\ref{fig:sor_orientation}).
Increasing $p_{SoR}$ from $0$ to $0.1$ improves cooperation rates by approximately 20--40\% if $p_{EV} > 0$.
However, further increases in $p_{SoR}$ beyond $0.2$ reduce cooperation rates.
This reduction occurs because excessive $p_{SoR}$ makes agent collisions more likely, which in turn hinders effective migration.
These findings suggest that some randomness in agent mobility is necessary to maintain a high level of cooperation.

\subsubsection{Payoff parameters and mutation rate}

We also examined the effects of varying the payoff parameters $(T, S)$ across several game structures, as well as the mutation rate $\mu$, including $\mu = 0$ and $\mu = 0.01$.
While these variations did not yield qualitatively new insights, they confirmed the robustness of the observed patterns.
The corresponding results are provided in the Supplementary Material for completeness.

\section{Discussion}\label{sec:discussion}

%================================
% Summary
%================================
We examine the joint effects of environmental variability and agent mobility on the evolution of cooperation to understand the causal dynamics among these factors in spatially structured populations.
Our model incorporates unpredictable environmental variability by implementing SoRs that move randomly across a two-dimensional space, generating dynamic spatial heterogeneity in resource availability.
Agents accumulate resources through cooperative or competitive interactions and those with lower resource level become more likely to migrate to neighboring cells and update their strategies.
Using this model, we reveal three key findings:
(i) with sufficient agent mobility, even modest environmental variability promotes cooperation, but further variability does not enhance cooperation;
(ii) with any level of environmental variability, agent mobility promotes cooperation; and
(iii) these effects occur because environmental variability disrupts a few large stable defector groups that form in resource-rich areas, and agent mobility enables the formation of numerous small cooperator groups on those sites.

%================================
% Significance
%================================
These findings contribute to our understanding of the evolution of cooperation in several important ways.
First, we have discovered the joint impact of environmental variability and agent mobility on cooperation, which previous studies \cite{Brockhurst2007, Assaf2013, Miller2015, Amirjanov2017, Stojkoski2021, Kroumi2021, Vainstein2007, Cong2012, Chen2012, He2020, Dhakal2020, Ren2021, Yang2023, Zhang2025} examined separately.
Second, in the VSH \cite{Potts1998, Potts2015, Potts2018, Lupien2021}, environmental variability has traditionally been understood as a selective pressure on individual cognitive abilities during the MSA in Africa \cite{Schuck-Paim2008, Sol2008, Sol2009}, implying a pathway from the variability to enhanced intelligence and subsequently to increased sociality.
However, we have shown that the variability can also affect sociality directly.
Specifically, agents in our model collectively form cooperative groups in response to environmental variability, even though they simply repeat interactions, migration, and strategy updating without becoming smarter.
This discovery enriches our understanding of whether intelligence or sociality came first in human evolution and of their complementary coevolution \cite{Inaba2025, Boyd2009, Henrich2021, Melis2023, Dunbar2024}.
Third, as our model and the mechanisms that we have discovered are highly general and abstract, they are not limited to our context.
With suitable adaptations, there may be potential for interdisciplinary applications in areas such as biology, ecology, and understanding modern social phenomena.

%================================
% Limitation & Future work
%================================
Despite these contributions, our study has several limitations that suggest directions for future research.
First, because our model aims to understand abstract macroscopic phenomena rather than to reproduce or predict specific real-world outcomes, it employs simplifications at various levels.
Future research could address the simplifications by exploring other scenarios, for example:
game structures such as public goods games \cite{Ren2021, Zhang2025, Kollock1998} rather than pairwise interactions;
migration based not only on resource levels but also on social relationships between agents \cite{Cong2012, Yang2023};
strategy sets beyond All-C and All-D \cite{Dugatkin1991}; and
adaptive strategy updating that varies learning mechanisms \cite{Ohtsuki2022, Turner2023} based on environmental conditions.
Second, this study will ultimately require a more formal mathematical foundation.
As is typical of multi-agent approaches, our model---though much simpler than the real world---remains too complex for analytical treatment.
Thus, we need to explore the prospects for mathematical formalization by reproducing key aspects of our results with simpler models.
Third, our research relies primarily on general patterns described in the VSH rather than on empirical data from the MSA in Africa.
This limitation is common of theoretical studies on the evolution of cooperation.
This is because evidence from this ancient period is sparse, and more fundamentally, cooperative behaviors cannot be directly preserved in archaeological remains, leaving only indirect evidence available.
Although these constraints are fundamental, we hope that theoretical studies like ours will play a crucial role in directing empirical research and providing frameworks for interpreting empirical findings.

\section*{Supporting information}

\subsection*{Data and code availability}
All data, along with the code needed to run the simulations and generate the figures, are available at \\ https://github.com/mas178/Inaba2025-2D.

\section*{Contributions}

M.I. was responsible for designing and implementing the research, building the model, analyzing the results, and writing the manuscript.
E.A. supervised the research.
All authors reviewed and approved the final manuscript.

\section*{Declaration of generative AI and AI-assisted technologies in the writing process}

During the preparation of this work, the authors used ChatGPT (OpenAI) and Claude (Anthropic) in order to improve the language and clarity of the manuscript.
After using these tools, the authors reviewed and edited the content as needed and take full responsibility for the content of the publication.

\bibliographystyle{main}
\bibliography{refs}

\end{document}